\documentclass[%
 reprint,
showpacs
 amsmath,amssymb,
prb,
]{revtex4-1}
\usepackage{amsmath}
\usepackage{graphicx}
\usepackage{dcolumn}
\usepackage{bm}

\begin{document}

\preprint{APS/123-QED}

\title{Applicability of point dipoles approximation to all-dielectric metamaterials}

\author{S.M. Kuznetsova}
\affiliation{Radiophysics Department, University of Nizhny Novgorod, Nizhny Novgorod 603950, Russia}
\author{A. Andryieuski}
\author{A.V. Lavrinenko}
\email{alav@fotonik.dtu.dk}
\affiliation{DTU Fotonik, Technical University of Denmark, DK-2800 Kongens Lyngby, Denmark}%





\date{\today}

\begin{abstract}
 
 All-dielectric metamaterials consisting of high-dielectric inclusions in a low-dielectric matrix are considered as a low-loss alternative to resonant metal-based metamaterials. In this contribution we investigate the applicability of the point electric and magnetic dipoles approximation to dielectric meta-atoms on the example of a dielectric ring metamaterial. Despite the large electrical size of high-dielectric meta-atoms, the dipole approximation allows for accurate prediction of the metamaterials properties for the rings with diameters up to $\approx0.8$ of the lattice constant. The results provide important guidelines for design and optimization of all-dielectric metamaterials.

\end{abstract}
\pacs{42.25.-p, 77.55.-g, 78.67.Pt}
\maketitle


\section{\label{I}Introduction}
In recent years metamaterials have aroused great attention as a tool for achieving desired electromagnetic properties, which might not exist in nature [\onlinecite{MM1}-\onlinecite{MM5}]. 
Among them a special attention has been paid to all-dielectric metamaterials [\onlinecite{DMM1}-\onlinecite{DMM3}], which do not suffer from the Ohmic losses contrary to the metal-based ones.
All-dielectric metamaterials with various constitutive units, such as spheres [\onlinecite{Vendik}], cubes [\onlinecite{Zhang}] and rods [\onlinecite{OBrien}, \onlinecite{Schuller}], were shown to exhibit the negative magentic permeability as well as negative [\onlinecite{Vendik1}-\onlinecite{JWang}] and zero refractive index [\onlinecite{Moitra}]. 
Silicon nanodimers, exhibiting both electric and magnetic dipole responses, have recently been experimentally demonstrated to enhance electric and magnetic fields in the optical range [\onlinecite{Bakker}]. 
High permittivity dielectric rings are of particular interest, because they possess a strong broadband magnetic response [\onlinecite{Jelinek}-\onlinecite{Bakunov}]. In addition, they have more degrees of freedom for tuning the resonance frequency by geometrical parameters than, for example, spheres, cubes or cylinders.
Furthermore, they exhibit resonances on the THz frequencies [\onlinecite{Maslov}-\onlinecite{Diel3}], a very promising range for emerging applications.

Being as the most frequent appearence periodical structures composed of subwavelength elements (the metamaterial lattice constant is on the order of $\lambda/20-\lambda/4$), three dimensional metamaterials are often treated as continuous media (the procedure referred to as \textit{homogenization}), described by some effective parameters, e.g. permittivity $\varepsilon$, permeability $\mu$, refractive index $n$ and impedance $z$, that simplifies their description. 
Various methods of homogenization have been suggested based on, for instance, field averaging, single-interface scattering, non-local dielectric function and other approaches (see the brief overview of the methods in [\onlinecite{Andr}]). 
The most popular practical way of retrieving the effective parameters remains the Nicolson-Ross-Weir (NRW) or S-parameters method based on inversion of formulas for Fresnel's reflection and transmission coefficients. 
It was originally proposed for conventional media [\onlinecite{Nic}] and then extended to metamaterials [\onlinecite{Smith1}].
The NRW method was accused of the loss of the retrieved effective parameters' physical meaning, since they do not satisfy the locality principle [\onlinecite{Sim0}-\onlinecite{Sim2}] and presence of the so-called "anti-resonance" [\onlinecite{Anti}] (non-physical for passive materials Im$(\varepsilon)<0$ and Im$(\mu)<0$; we assume optical exponential convention $\exp(-i\omega t)$).
Moreover, introduction of bulk effective parameters to metasurfaces or few layers metamaterials is questionable.
Obviously the properties of a single layer or a layer with only few neighbors can be substantially different from the properties of a layer with many neighbors.
In some cases of strong interaction between meta-atoms it is not possible to introduce any meaningful effective parameters [\onlinecite{Andr1}].

We use another approach in which meta-atoms are replaced with point electric and magnetic dipoles [\onlinecite{Sim1}, \onlinecite{SimTr}, \onlinecite{Albo}-\onlinecite{Sim3}]. 
The advantage of such approach is that the interaction of dipoles can be explicitly taken into account either numerically for the arrays up to few hundreds by few hundreds elements or analytically for an infinite array [\onlinecite{Tretyakov}].

However, the applicability of the point dipoles approximation to all-dielectric metamaterials has not been verified yet.
In order to achieve strong resonant features, high refractive index dielectrics are typically used, and thus even geometrically subwavelength meta-atoms may become electrically (optically) larger than a wavelength exhibiting rapid field variations within their volume.

The goal of this work is to verify the applicability of the electric and magnetic dipole approximation for description of all-dielectric metamaterials. 
We make the study on the example of dielectric ring metamaterials.
We start with considering a single layer of meta-atoms and investigate the dependence of their electric and magnetic polarizabilities of the geometrical parameters. 
Then in order to figure out the limits of the dipole approximation we show how the retrieved polarizabilities depend on the metamaterial lattice constant.
Finally we consider a few layers of dielectric rings and compare the analytical predictions for reflectance $R$ with the full-wave numerical simulations. 

The paper is organized as follows. 
Sec.~\ref{II} is devoted to research methodology. 
The results of numerical simulations and its comparison with analytical predictions are presented in Sec.~\ref{III}. 
Discussion and conclusions are assembled in Sec.~\ref{IV}.

\section{\label{II} Methods}

\begin{figure*}[h!]
\includegraphics{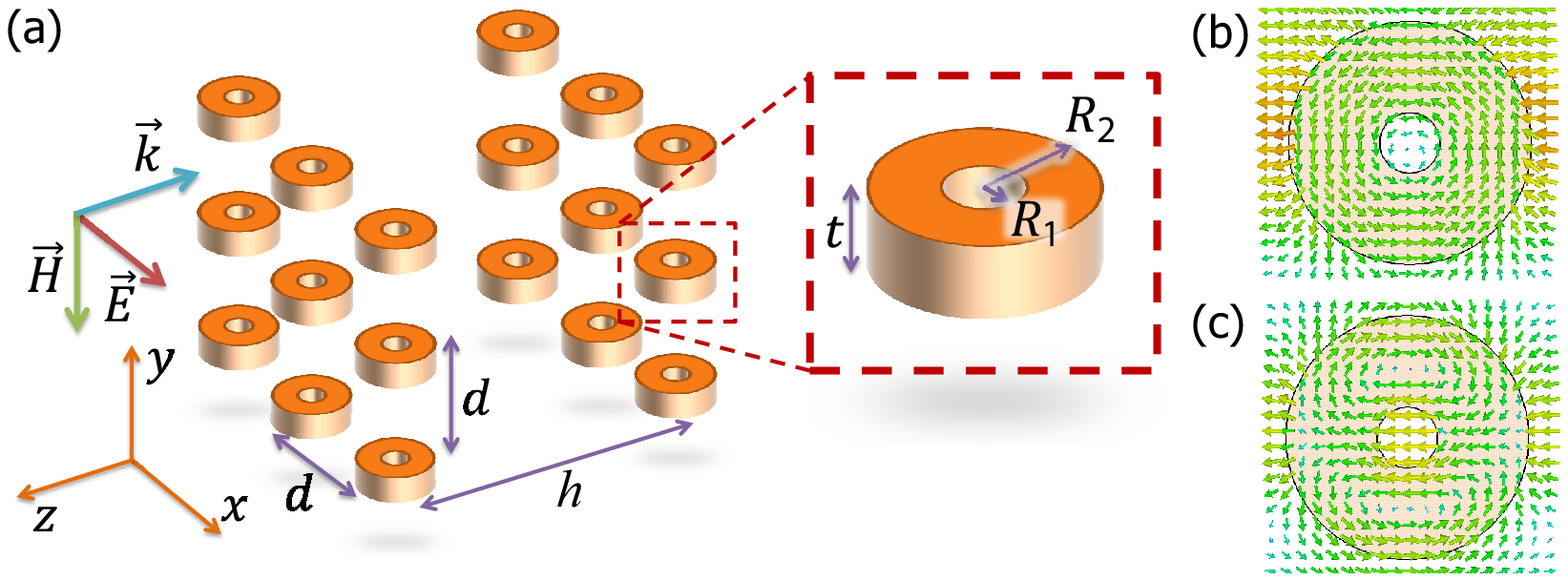}
\caption{\label{fig:1} (Color online). (a) Dielectric ring metamaterial consists of rings arranged in a square lattice. The plane wave is incident perpendicularly to the layers. Electric field distribution in the unit cell for magnetic (b) and electric (c) dipole resonances excited with the incident plane wave.}
\end{figure*}

We consider dielectric rings of the outer and inner radii $R_2$ and $R_1$, respectively, and thickness $t$ arranged in the square lattice with lattice constant $d$ within each layer and distance between the layers $h$ [see the geometrical parameters in Fig. ~\ref{fig:1}(a)]. 
The rings consist of dielectric with material permittivity $\varepsilon=100+i$ suspended in free space. 
A plane wave is normally incident along the $z-$axis and its electric field directed along the $x-$axis.
We are interested in the lowest order resonances of the rings classified as magnetic and electric dipole modes [see Fig. ~\ref{fig:1}(b) and (c)]. 

In order to obtain electric and magnetic dipole polarizabilities $\tilde{\alpha}_e$ and $\tilde{\alpha}_m$ for an individual meta-atom, we treat every ring in the array as a combination of point electric and magnetic dipoles [\onlinecite{Albo}] oriented along the $x-$ and $y-$ axes, respectively, in correspondence with the polarization of the incident wave. 
The equivalent dipoles are positioned in the center of the rings. 
The electric $\tilde{p}$ and magnetic $\tilde{m}$ dipole moments are induced by the external field (amplitude of the electric and magnetic field are $E_0$ and $H_0$, respectively and $E_0=\eta_0H_0$, where $\eta_0=\sqrt{\mu_0/\varepsilon_0}=120\pi [\Omega]$ is the free-space impedance) as well as by  all the neighbors within the infinite array (interaction field $E_{inter}, H_{inter}$). 
In the case of a single layer, taking into account the identity of the dipoles, we express $\tilde{p}$ and $\tilde{m}$
\begin{eqnarray}
\tilde{p}=\tilde{\alpha}_e(E_0+E_{inter})=\tilde{\alpha}_e(E_0+\varepsilon_0 d^3\beta^0\tilde{p}),\\
\tilde{m}=\tilde{\alpha}_m(H_0+H_{inter})=\tilde{\alpha}_m(H_0+\mu_0 d^3\beta^0\tilde{m}),
\end{eqnarray}

\noindent where the interaction constant $\beta^0$ is 
\begin{equation}\label{eq:beta0}
\beta^0 = Re\left[\frac{ikd}{4}\left(1+\frac{1}{ikR_0}\right)e^{ikR_0}\right]+i\left(\frac{kd}{2}-\frac{k^3d^3}{6\pi}\right),
\end{equation} 
\noindent  and $R_0$=d/1.438 [\onlinecite{Tretyakov}].

The amplitude reflection and transmission coefficients can be split into electric $R_e$ and magnetic $R_m$ terms[\onlinecite{Albo}]:
\begin{eqnarray}
R=R_e+R_m, \quad T=1+R_e-R_m,
\\
R_e=\frac{i k d}{2} \frac{1}{\frac{\varepsilon_0 d^3}{\tilde{\alpha}_e}-\beta^0}, \quad R_m=-\frac{i k d}{2} \frac{1}{\frac{\mu_0 d^3} {\tilde{\alpha}_m}-\beta^0}. 
\end{eqnarray}
The polarizabilities of the particles are then expressed as
\begin{equation}
\tilde{\alpha}_e=\frac{\varepsilon_0 d^3}{\beta^0+\frac{ik d}{2R_e}}, \quad 
\tilde{\alpha}_m=\frac{\mu_0 d^3}{\beta^0-\frac{ik d}{2R_m}}.
\end{equation}

We introduce the normalized dimensionless polarizabilities $\alpha_e=\tilde{\alpha}_e/\varepsilon_0 d^3$, $\alpha_m=\tilde{\alpha}_m/\mu_0 d^3$. Then the inverse polarizabilities are
\begin{equation}
\frac{1}{\alpha_e}=\beta^0+\frac{ikd}{R+T-1}, \quad 
\frac{1}{\alpha_m}=\beta^0-\frac{ikd}{R-T+1}.
\end{equation}

Considering two layers of particles one should take into account the interaction between the layers. The wave propagates as shown in Fig.1(a), and the coordinate system origin is placed in the center of a ring in the first layer. 
The propagation occurs towards negative $z$, $E(z)=E_0exp(-ikz)$.
The system of equations [\onlinecite{Tretyakov}] for the normalized dimensionless dipole moments $p_{1,2}=\tilde{p}_{1,2}/(\varepsilon_0d^3E_0), m_{1,2}=\tilde{m}_{1,2}/(\mu_0d^3H_0)$ reads
\begin{eqnarray}
\frac{1}{\alpha_e}p_1=1+\beta^0 p_1+\beta^{-h} p_2+\beta_{em}^{-h} m_2,
\\
\frac{1}{\alpha_e}p_2=exp(ikh)+\beta^0 p_2+\beta^{h} p_1+\beta_{em}^{h} m_1,
\\
\frac{1}{\alpha_m}m_1=1+\beta^0 m_1+\beta^{-h} m_2+\beta_{me}^{-h} p_2,
\\
\frac{1}{\alpha_m}m_2=exp(ikh)+\beta^0 m_2+\beta^{h} m_1+\beta_{me}^{h} p_1,
\end{eqnarray}
\noindent where the interaction coefficients can be found with approximate formulas (valid up to $kd\lesssim1.5$)
\begin{eqnarray}\label{eq:betah}
\beta^h=-Re\left\lbrace \frac{-ikd}{4}\left[\left(1+\frac{1}{ik\sqrt{R_0^2+h^2}}\right)+\nonumber \right.\right.\\
\left.\left.+\frac{h^2}{R_0^2+h^2}\left(1-\frac{1}{ik\sqrt{R_0^2+h^2}}\right)\right]e^{ik\sqrt{R_0^2+h^2}}+\right.\nonumber\\
\left.+\frac{d^3}{4\pi}\left(\frac{1}{h^3}-\frac{ik}{h^2}-\frac{k^2}{h}\right)e^{ikh}\right\rbrace+i\frac{kd}{2}cos(kh),\nonumber
\\
\beta_{em}^h=-i\frac{kd}{2}\frac{h}{\sqrt{R_0^2+h^2}}e^{ik\sqrt{R_0^2+h^2}},
\end{eqnarray}
and the following equalities are valid
\begin{equation}
\beta^{-h}=\beta^h, \beta_{em}^h=-\beta_{me}^h=-\beta_{em}^{-h}.
\end{equation}

Thus for two meta-layers we get the system of linear equations for unknown dipole moments:
\begin{equation}
-\begin{bmatrix}
\beta^0-\frac{1}{\alpha_e} & \beta^h & 0 & \beta_{em}^h \\
\beta^h & \beta^0-\frac{1}{\alpha_e} & -\beta_{em}^h & 0\\
0 & \beta_{em}^h & \beta^0-\frac{1}{\alpha_m} & \beta^h\\
-\beta_{em}^h & 0 & \beta^h & \beta^0-\frac{1}{\alpha_m}
\end{bmatrix}
\begin{bmatrix}
p_1\\
p_2\\
m_1\\
m_2
\end{bmatrix}= 
\begin{bmatrix}
1\\
e^{ikh}\\
1\\
e^{ikh}
\end{bmatrix}.
\end{equation}

After finding the dipole moments one can calculate the electric and magnetic contributions to reflection and transmission coefficients
\begin{eqnarray}
R_e=\frac{1}{2}ikd(p_1+p_2 e^{ikh}),
\\
R_m=-\frac{1}{2}ikd(m_1+m_2 e^{ikh}),
\\
T_e=\frac{1}{2}ikd(p_1 e^{ikh}+p_2),
\\
T_m=\frac{1}{2}ikd(m_1 e^{ikh}+m_2),
\end{eqnarray}
and finally obtain R and T
\begin{eqnarray}
R=\frac{1}{2}ikd(p_1+p_2 e^{ikh}-m_1-m_2 e^{ikh}),
\\
T=e^{ikh}+\frac{1}{2}ikd(p_1 e^{ikh}+p_2+m_1 e^{ikh}+m_2).
\end{eqnarray}

For larger number $N$ of layers the procedure is analogical. At first one finds the dipole moments solving the system of equations
\begin{widetext}
\begin{eqnarray}
-\begin{bmatrix}
\beta^0-\frac{1}{\alpha_e} & \beta^h & \beta^{2h} & \vdots & \beta_{em}^{(N-1)h}\\
\beta^h & \beta^0-\frac{1}{\alpha_e} & \beta^h & \vdots & \beta_{em}^{(N-2)h}\\
\beta^{2h} & \beta^h & \beta^0-\frac{1}{\alpha_e} & \vdots & \beta_{em}^{(N-3)h}\\
\dots & \dots & \dots & \ddots & \dots & \dots\\
-\beta_{em}^{(N-2)h} & -\beta_{em}^{(N-3)h} & -\beta_{em}^{(N-4)h} & \vdots & \beta^{h}\\
-\beta_{em}^{(N-1)h} & -\beta_{em}^{(N-2)h} & -\beta_{em}^{(N-3)h} & \vdots & \beta^0-\frac{1}{\alpha_m}
\end{bmatrix}
\centerdot
\begin{bmatrix}
p_1\\
p_2\\
p_3\\
\vdots\\
m_{N-1}\\
m_N
\end{bmatrix}
=
\begin{bmatrix}
1\\
e^{ikh}\\
e^{i2kh}\\
\vdots\\
e^{ikh(N-2)}\\
e^{ikh(N-1)}
\end{bmatrix},
\end{eqnarray}
\end{widetext}

\noindent and then calculates the reflection and transmission coefficients
\begin{eqnarray}
R=\frac{1}{2}ikd \sum_{n=1}^{N} (p_n-m_n) e^{ikh(n-1)},\\
T=e^{ikh(N-1)}+\frac{1}{2}ikd \sum_{n=1}^{N} (p_n+m_n) e^{ikh(N-n)}.
\end{eqnarray}

We performed numerical full-wave simulations with CST Microwave Studio [\onlinecite{CST}] in the time domain with effectively periodic (perfect electric along $x-$ and perfect magnetic along $y-$axis) boundary conditions and normally incident plane wave excitation.

\section{\label{III} Results}

For the rest of the paper we consider the thickness of the rings fixed to $t/d=0.2$, distance between the centers of the rings in neighbor layers $h=d$ and variable inner and outer radii $R_1, R_2$. The frequency range is 0-0.3 $d/\lambda$ in order to stay within the metamaterial regime. 
To be noted, formulas (\ref{eq:beta0}),(\ref{eq:betah}) for the interaction coefficients are applicable for $kd\lesssim1.5$ [\onlinecite{Tretyakov}], which corresponds to $d/\lambda<0.25$.

\subsection{Variation of the outer $R_2$ and inner $R_1$ radii}
\begin{figure}[h!]
    \includegraphics{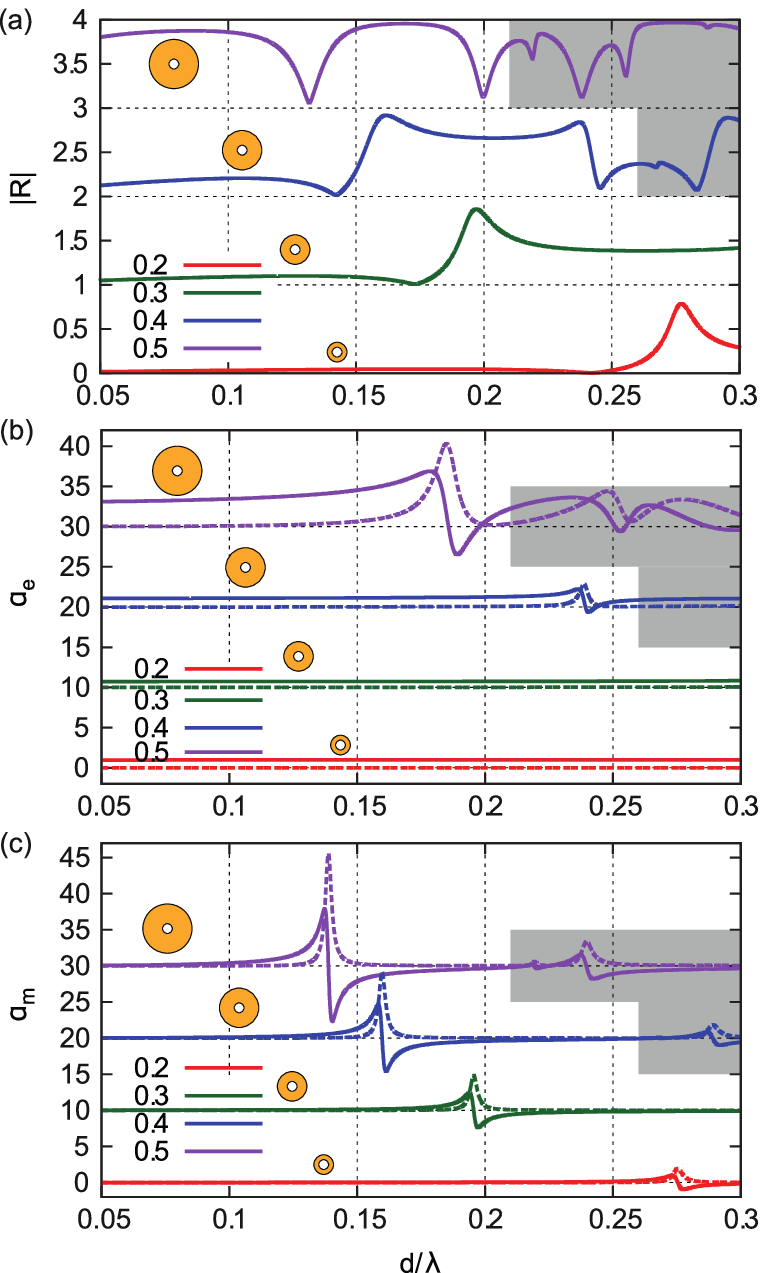}
\caption{\label{fig:2}(Color online). (a) Absolute value of the reflection $R$ for a single array of rings. Dimensionless electric (b) and magnetic (c) polarizabilities. The inner radius is fixed to $R_1/d=0.1$. The outer radius $R_2/d=$ 0.2 (red), 0.3 (green), 0.4 (blue) and 0.5 (purple). For clarity of presentation a vertical incremental offset is introduced to the spectra. The offsets of the spectra are marked with horizontal dotted line. This convention is kept throughout the rest of the article. The grey-shaded areas mark the range where higher-order modes are excited.} 
\end{figure}

In this section we investigate the influence of geometrical parameters, namely, the outer and inner radii on the resonant properties of the dielectric ring metamaterial.
Let us fix the inner radius $R_1/d=0.1$ and vary the outer radius in the range $R_2/d=0.2-0.5$ (Fig.\ref{fig:2}).
The resonant peaks and dips are clearly seen in the reflectance spectra, for example, follow the blue line in Fig. \ref{fig:2}(a). 
They correspond to the lower-order modes of the dielectric ring resonators, namely, the magnetic dipolar (lower frequency) and electric dipolar (higher frequency) modes. The corresponding electric field distributions are shown in Fig. \ref{fig:1}(b) and (c), respectively.
Consequently, the retrieved electric [Fig. \ref{fig:2}(b)] and magnetic [Fig. \ref{fig:2}(c)] polarizabilities exhibit resonance behavior at the frequencies of the reflection maxima.
Increase of the size of the rings expectedly leads to decrease in the resonant frequency and increase of the resonance strength [Fig.\ref{fig:2}(b),(c)].
Apart from the low-order dipolar modes some higher order modes can be excited (see shaded area in Fig. \ref{fig:2}), however, they are lying in the high frequency zone and will be excluded from the further analysis. 

\begin{figure}[h!tb]
    \includegraphics{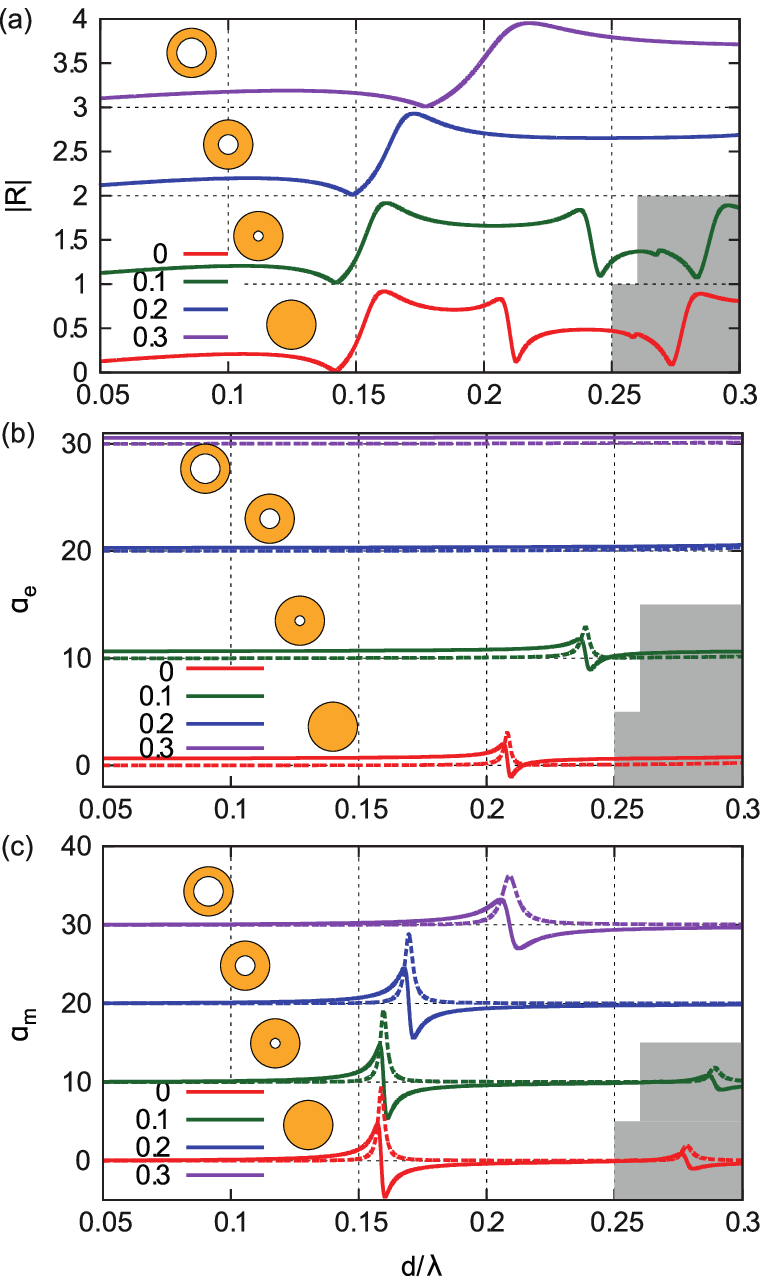}
\caption{\label{fig:3}(Color online). (a) Reflection $R$ for a single array of rings and their dimensionless electric (b) and magnetic (c) polarizabilities.  The outer radius is fixed $R_2/d=0.4$. The inner radius varies $R_1/d=$ 0 (red), 0.1 (green), 0.2 (blue) and 0.3 (purple).}
\end{figure}

When $R_2$ is fixed, but inner radius $R_1$ is varied, one of the reflection peaks stays almost unchanged whereas  another one shifts dramatically [Fig. \ref{fig:3}(a)]. 
Increasing $R_1$ influences strongly on the location of the electric dipole peak [Fig. \ref{fig:3}(b)], but has weaker impact on the magnetic dipole [Fig. \ref{fig:3} (c)].
The magnetic field of the excited resonance is induced by the circulation of the displacement current. Accordingly to the mode structure [Fig.1(b)], it is obvious that the dominating contribution is given by the periphery of the ring.
Overall the modes blue-shift with the inner hole radius $R_1$ growth [Fig. \ref{fig:3}(b) and (c)] as the result of the decreasing high-dielectric material filling fraction.

\subsection{Applicability of the dipole approximation and variation of the array period}

For retrieval of ring electric $\alpha_e$ and magnetic $\alpha_m$ polarizabilities, which are the characteristics of a single ring, we use formula (7) and simulate reflection and transmission for an infinite periodic array of the rings.
The influence of the infinite number of neighbors is taken into account through interaction constant $\beta^0$, for which accurate analytical formulas exist for the square array of point dipoles [\onlinecite{Tretyakov}].
At the same time, the rings we consider here are not point-like, but rather comparable to the period of the array.
This may lead to both deviation of the field distribution of a certain dipole from the field of the dipole located in its center (thus, influence on polarizabilities $\alpha_e$ and $\alpha_m$) as well as on the interaction field of all other dipoles (thus, influence on interaction constant $\beta^0$).
We assume that interaction constant $\beta^0$ is the same as given by formula (7), and thus all effects related to the finite size of the rings are included into retrieved polarizabilities $\alpha_e$ and $\alpha_m$.
Then, retrieving the polarizabilities for the arrays of various periods we are able to characterize how different from the point dipoles the meta-atoms are (ideally the retrieved polarizabilities should not depend on the array period).
We call the actual period of the structure $d^*$ in order to differ it from the previously used constant $d$, while the latter is now used for normalization of frequency and polarizabilities only.

\begin{figure*}[h!]
    \includegraphics{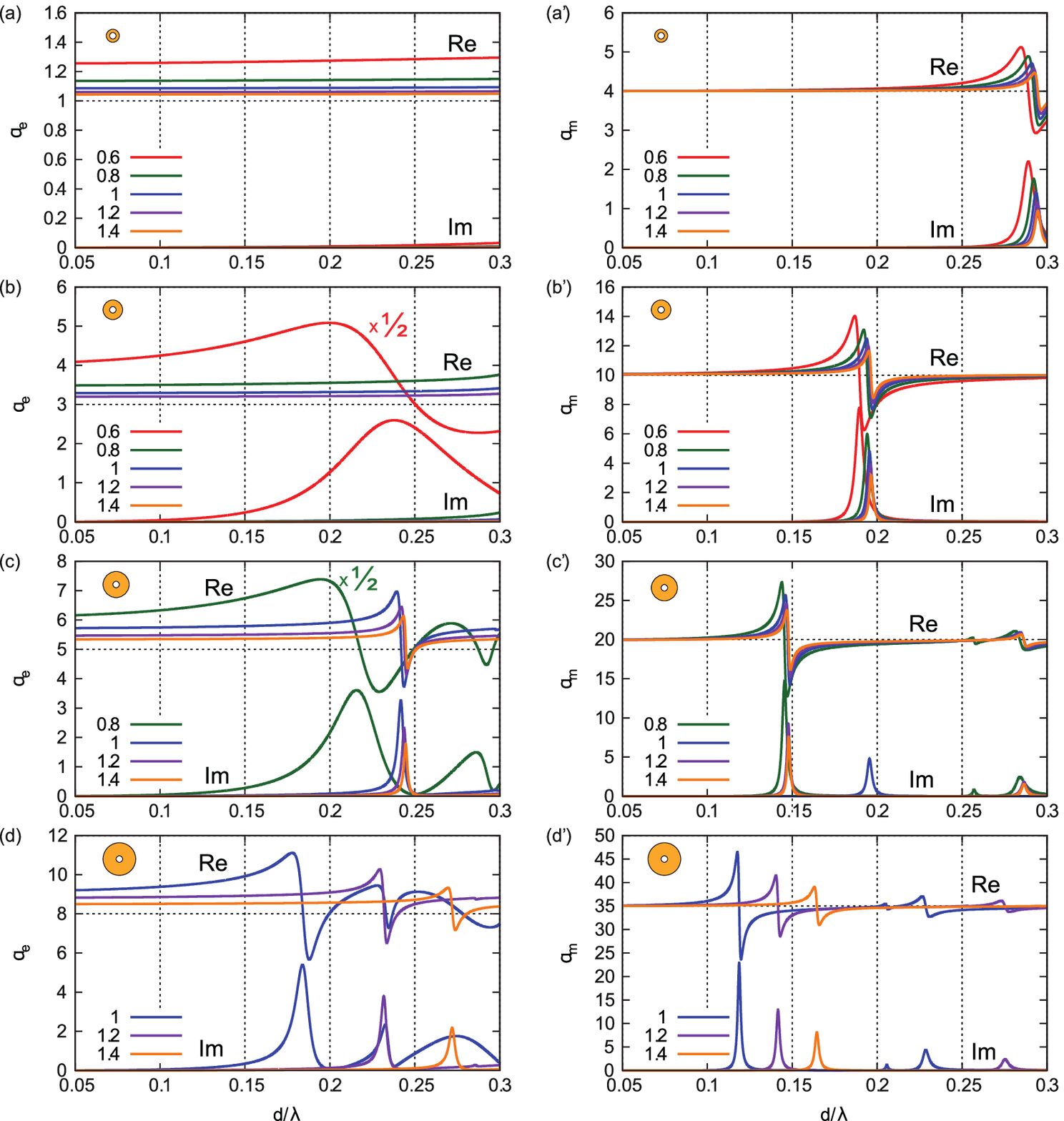}
\caption{\label{fig:4} (Color online). Normalized electric (a)-(d) and magnetic (a')-(d') polarizabilities for $R_2/d$ = 0.2, 0.3, 0.4 and 0.5, respectively. The values of the actual period $d^*/d$ are (not on all graphs, only the cases of $d^*\geq 2R_2$ are shown) 0.6 (red), 0.8 (green), 1 (blue), 1.2 (purple) and 1.4 (orange). The real parts of the polarizabilities have positive offset for presentation clarity. The curves for $d^*/d=0.6$ (red) in (b) and $d^*/d=0.8$ (green) in (c) are multipled by $1/2$ in order to fit the graph.}
\end{figure*}

Period $d^*$ influences the restored electric and magnetic polarizabilities (Fig. \ref{fig:4}) and they converge to a certain value with increasing period $d^*$.
Such behavior is natural, since the larger the period the smaller is the radius-to-period ratio $R_2/d^*$ and, thus, the accuracy of the dipole approximation improves.
Generally convergence of magnetic polarizabilities $\alpha_m$ [Fig. \ref{fig:4}(a')-(d')] is faster than electric $\alpha_e$ [Fig. \ref{fig:4}(a)-(d)] what can mean that magnetic interaction of the rings is weaker than electric.
In case of touching rings ($d^*=2R_2$) the retrieved polarizabilities differ much from the asymptotic value, clearly exhibiting failure of the point dipoles approximation. 
Indeed, touching rings can hardly be considered as a set of separate rings, but rather as a periodic subwavelength grating or a perforated dielectric membrane.
The polarizabilities convergence is good for the rings radii up to $R_2/d=0.4$ [Fig. \ref{fig:4}(a)-(c) and (a')-(c')].
For larger rings with $R_2/d=0.5$ no convergence is observed for the period $d^*$ up to $d^*/d=1.4$ that lies on the boundary of the metamaterials regime and corresponds to $d^*\approx0.5\lambda$ at the normalized frequency $d/\lambda=0.3$ [Fig. \ref{fig:4}(d) and (d')].
Overall we may conclude that dipole approximation is consistent for the rings radius up to $R_2/d=0.4$ and lattice constant $d^*/d>0.8$.

\subsection{Several layers structure}
\begin{figure}[h!]
    \includegraphics{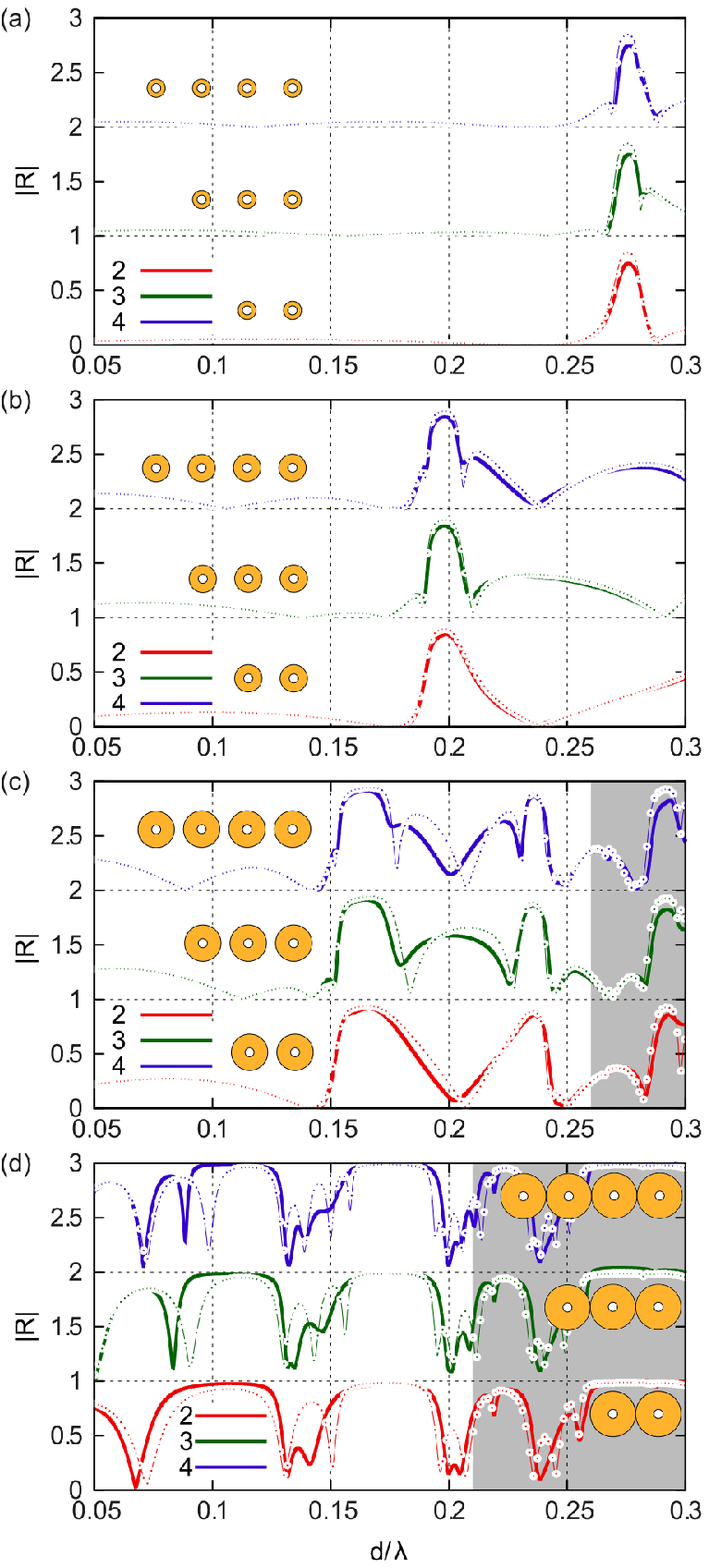}
\caption{\label{fig:5} (Color online). Absolute value of reflection $|R|$ for 2 (red), 3 (green) and 4 (blue) layers of rings. Analytical predictions (thick solid lines) are compared to full-wave numerical simulations (thin lines with circles). The inner radius is fixed $R_1/d=0.1$. The outer radius is $R_2/d=$ 0.2 (a), 0.3 (b), 0.4 (c) and 0.5 (d). Grey area corresponds to the frequency range where high-order resonances are excited.} 
\end{figure}

Knowing polarizabilities $\alpha_e$ and $\alpha_m$ of individual rings retrieved from the simulated reflection and transmission for a single layer, we may analytically predict the reflection and transmission for several layers of rings (see the methodology section for details).
Direct comparison of these analytical predictions to numerical simulations can give us additional evidences of applicability or non-applicability of the point dipoles approximation.
We use the same rings geometrical parameters as in the previous subsection.
Fig. \ref{fig:5} shows the reflection for different $R_2$ for 2, 3 and 4 layers structure with $R_1/d=0.1$ and $R_2/d=0.2, 0.3, 0.4$ and $0.5$.

For smaller radii $R_2/d=0.2$ and $0.3$ and even for $R_2/d=0.4$ the analytical prediction is in a good correspondence with the numerical simulations providing another confirmation that the dipole approximation works well [Fig. \ref{fig:5}(a)-(c)].
The discrepancy between the predicted and simulated results increases with $R_2$ and for the case of touching rings $R_2/d=0.5$ [Fig. \ref{fig:5}(d)] the reflections given with two methods do not resemble each other reasonably, especially in the region of resonant peaks and dips.
Interestingly, the reflection is different not only in the region of polarizabilities resonances (see Fig. \ref{fig:2}), but also in other regions of reflection dips which we interpret as appeared due to Fabry-Perot resonances (increasing $R_2$ leads to the high-dielectric materials filling fraction increase and thus to the increase of the overall optical thickness of the metamaterial slab).

We may conclude that the electric and magnetic polarizabilities restored from a plane square-lattice array are able to predict the several layers reflectance for $R_2/d<0.4$, but fail for larger radius. 

\section{\label{IV}Discussion  and conclusions}

We have analyzed electric and magnetic polarizabilities of the high-dielectric rings  approximating them as point dipoles.
The geometical parameters and the free-space wavelength are normalized to the unit cell size, thus the results of our analysis are wavelength-independent and virtually are applicable to any freqeuncy range.

Variations of geometrical parameters allow for tuning the resonance frequencies for $\alpha_e$ and $\alpha_m$ nearly independently: outer radius $R_2$ influences the resonance frequencies of both electric and magnetic polarizabilities (Fig. \ref{fig:2}) while the inner radius predominantly influences electric polarizability $\alpha_e$ (Fig. \ref{fig:3}).

We come to a natural conclusion that decreasing the  structural unit size leads to a better prediction of the electromagnetic properties with the point dipoles approximation.
The dipole approximation works well for the rings up to $R_2/d=0.4$ (Fig. \ref{fig:4}). 
This value is surprisingly large, since the electrical size of the ring at the frequency $d/\lambda=0.15$ is $2R_2n\approx1.2\lambda$, thus the ring is not electrically small anymore and the field varies rapidly in space within the ring.

For larger rings the situation is different.
The rings with the radius $R_2/d=0.5$ are not isolated -- they are literally touching each other. 
In this case the dipole approximation is not valid anymore. 
The retrieved polarizability changes dramatically with the lattice constant of the array [Fig. \ref{fig:4} (d)-(d')] and cannot predict the electromagnetic response for several layers [Fig. \ref{fig:5}(d)]. 
Instead of isolated particles separated by the free-space gaps, we rather get a perforated high-index dielectric grating. 
Even though the period of the grating is smaller than the free-space wavelength, and thus no diffraction to the free space can happen, the guided mode in such dielectric membranes can be excited. 
The power coupled at some point is no longer localized within the particle, but can propagate within the dielectric layer, thus the metamaterial cannot be effectively homogenized. 

We have investigated the metamaterial made of high-dielectric rings. 
Even though the selected $\varepsilon=100$ is modest (dielectric constant of barium strontium titanate may reach several hundreds [\onlinecite{BST}]), it is representative enough to make general enough conclusions. 
We may accurately anticipate the behaviour of the lower-index dielectrics. 
They might better suite to homogenization, having smaller refractive index contrast with the background. 
However, the unusual properties, based on strong electric and magnetic resonances, would also be weaker.

In conclusion, the all-dielectric ring metamaterials can be described with electric and magnetic point dipole approximation if the meta-atoms size (ring radius) is smaller than 0.8 of the period. 
For the meta-atoms that almost touch each other, the dipole approximation is not applicable.
We believe, the presented methodology and analysis results will be useful for development of low-loss all-dielectric based metamaterials.

\end{document}